\documentclass[11pt]{article}
\usepackage{psfig}
\newcommand{\bvec}[1]{\mbox{\boldmath ${#1}$}}

\title{\bf \large HYPERNUCLEAR PHYSICS WITH PHOTONS\footnote{Keynote
talk at the JLab workshop on Hypernuclear
Physics with Electromagnetic Probes (HYPLAB99), Jefferson Lab,
Dec. 2-4, 1999} }
\author{\normalsize C. Bennhold$^{\rm \;a,b}$,
        T. Mart$^{\rm \;c,b}$,
        F.X. Lee$^{\rm \;a}$,
        H. Haberzettl$^{\rm \;a}$,
        H. Yamamura$^{\rm \;d}$,
        K. Miyagawa$^{\rm \;d}$,\\
        W. Gl\"{o}ckle$^{\rm \;e}$,
        S.S. Kamalov$^{\rm \;f,b}$,
        L. Tiator$^{\rm \;b}$,
        L.E. Wright$^{\rm \;g}$}
\date{}

\begin{document}

\maketitle

\noindent $^{\rm a}\;${\small \it Center for Nuclear Studies,
        Department of Physics, George Washington
        University, Washington, }\\
{\small\it $~~~~$D.C. 20052, USA}

\noindent $^{\rm b}\;${\small\it Institut f\"ur Kernphysik,
 Johannes Gutenberg-Universit\"at, 55099 Mainz, Germany}

\noindent $^{\rm c}\;${\small\it Jurusan Fisika, FMIPA,
        Universitas Indonesia, Depok 16424, Indonesia}

\noindent $^{\rm d}\;${\small\it Department of Applied Physics,
        Okayama University of Science 1-1 Ridai-cho, Okayama 700, Japan}

\noindent $^{\rm e}\;${\small\it Institut f\"ur Theoretische
             Physik II, Ruhr-Universit\"at Bochum, D-44780 Bochum, Germany}

\noindent $^{\rm f}\;${\small\it Laboratory of Theoretical
            Physics, JINR Dubna, SU-101000, Moscow, Russia}

\noindent $^{\rm g}\;${\small\it Department of Physics, Ohio
            University, Athens, Ohio 45701, USA}

\vspace{5mm}

\begin{abstract}
We review the opportunities and challenges in the field of
hypernuclear physics with electromagnetic probes.  An overview is
presented regarding our current understanding of the elementary
production process on the nucleon.  This amplitude is then used in
the nuclear environment to study the hyperon-nucleon ($YN$)
interaction.  We discuss two scenarios: hypernuclear excitation
that allows the investigation of hypernuclear structure and the
bound $\Lambda$ in the nucleus, and quasifree kaon production on
the deuteron and on nuclei, which permits a more direct access to
the $YN$ force. Specific examples are given for few-body systems
and shell-model nuclei.
\end{abstract}

\section{\bf \normalsize ELEMENTARY KAON PHOTOPRODUCTION ON THE NUCLEON}

Since we are still a long way from calculating the scattering and
electromagnetic production of mesons on baryons directly from QCD,
effective field-theoretical descriptions in terms of purely
hadronic degrees of freedom are usually employed to compute such
processes.  QCD is assumed to provide the justification for the
parameters or cutoff functions used in the various approaches. At
threshold, SU(2) Chiral Perturbation Theory has been moderately
successful to describe pion photo- and electroproduction; attempts
to expand these techniques into the SU(3) arena to eta and kaon
production are still in their infancy.  In the resonance region
one usually relies on effective Lagrangian approaches where a
potential or driving term is defined that includes standard
nonresonant $s$-,$t$-, and $u$-channel poles along with resonances in
each of these channels as bare fields which are then dressed
through the final-state interaction. Only the dressed $s$-channel
poles above threshold are identified with physical resonances, the
nonpolar part is considered background for the particular process
(even though it may contain bare baryon resonances in the
$u$-channel).

While dynamical models involving various approximations for the
Bethe-Salpeter equation are becoming increasingly successful in
the description of pion photoproduction, the hadronic final state
interaction in kaon photoproduction has usually been neglected.
Without rescattering contributions the $T$-matrix is simply
approximated by the driving term alone which is assumed to be
given by a series of tree-level
 diagrams \cite{saghai96,williams92,mart95,han99,bennhold89,rosenthal88}.
Clearly, neglecting the final meson-baryon interaction in the full
meson photoproduction $T$-matrix automatically leads to violation
of unitarity since flux that can "leak out" into inelastic
channels has not been properly accounted for.  Enforcing unitarity
dynamically requires solving a system of coupled channels with all
possible final states.

The most recent coupled-channels approach within an effective
Lagrangian framework has been developed by Feuster and Mosel
\cite{feuster98,feuster99} and extended to higher energies and
additional channels by Waluyo {\it et al}. \cite{waluyo2000}.
Nucleon resonance parameters are extracted by simultaneously
analyzing all available data for reactions involving the initial
and final states $\gamma N, \pi N, \pi \pi N, \eta N, K \Lambda, K
\Sigma$ and $\eta' N$ up to $W = 2.0$ GeV.  The calculations
employ the $K$-matrix approximation, placing both intermediate
particles on their mass shell.  This procedure still allows for
the resonance widths to be generated dynamically, while the real
part of the self-energy is absorbed in an effective resonance mass
that is determined by the fit.

\subsection{The Born terms: SU(3) coupling constants, form factors and gauge invariance}

Our understanding of the kaon-baryon interaction is still much
poorer than our knowledge of the pion-nucleon force, exemplified
by the uncertainty in the kaon-hyperon-nucleon coupling constants
$g_{K\Lambda N}$ and $g_{K\Sigma N}$. Unlike the well established
pion-nucleon interaction which yields a pion-nucleon coupling
constant $g_{\pi NN}^2/4\pi$ around 13.7, the kaon coupling
constants extracted from different reactions (from hadronic to
electromagnetic) have much larger uncertainties, as shown in
Table\,\ref{tab:cct}. Most isobar models for kaon photoproduction
over the last 30 years have left the leading $KYN$ couplings as
open parameters to be determined by the data. Constraining these
values to within the SU(3) range gave results which were
overpredicting the data by up to a factor of 10.  Therefore, when
left as free parameters the couplings came out to be significantly
smaller than the SU(3) range. On the other side, most extractions
based on hadronic reactions yielded couplings constants well
within the SU(3) limits.  This discrepancy suggested that an
important piece of physics has been left out in isobar models: the
extended structure of the hadrons, parametrized in terms of a
hadronic form factor.

\begin{table}[!t]
\begin{center}
\caption{The Born coupling constants obtained from various
sources. } \label{tab:cct}
\begin{tabular}{lccc}
\hline\hline\\ Source & ${\displaystyle \frac{g_{K\Lambda
N}}{\sqrt{4\pi}}}$& ${\displaystyle \frac{g_{K\Sigma
N}}{\sqrt{4\pi}}}$& Reference\\ [2.5ex] \hline\\ SU(3)&$-4.40$ to
$-3.0$&$+0.9$ to $+1.3$&\cite{adel2}\\ $K$-$N$
scattering&$~|3.53|$&$~|1.53|$&\cite{anto}\\ $Y$-$N$
scattering&$-3.86$&$+1.09$&\cite{timmer}\\ $N{\bar N}\rightarrow
Y{\bar Y}$ LEAR data & $-3.92$ & - & \cite{timmer}\\ QCD sum
rules&$-2.82$ to $-1.96$&$0.25$ to
$0.80$&\cite{choe96,aliev2000}\\ $\gamma p \rightarrow K^+
\Lambda(\Sigma)$ (extrapolation) & $|3.52|$&$ - $&\cite{deo74}\\
$\gamma p \rightarrow K^+ \Lambda(\Sigma)$ (isobar models) &
~~~~$-4.2$ to $-0.9$~~~~ & $+0.02$ to $1.8$& ~~~many~~~\\ \\

\hline\hline
\end{tabular}
\end{center}
\end{table}

However, it is well-known that the sum of the first three
photoproduction diagrams---i.e., the sum of the $s$-, $u$-, and
$t$-channel diagrams---is gauge-invariant only for bare hadronic
vertices with pure pseudoscalar coupling. Thus, for this most
basic case, the addition of a fourth contact-type graph is not
necessary for preserving gauge invariance. In all other instances,
however, one needs additional currents to ensure gauge invariance
and thus current conservation. For bare hadronic vertices with
pseudovector coupling, this extra current is the well-known
Kroll-Ruderman contact term.

Irrespective of the coupling type, however, most isobaric models
with bare vertices show a divergence at higher energies, which
clearly points to the need for introducing hadronic form factors
to cut off this undesirable behavior. For example, recent
calculations \cite{saghai96,mart95} demonstrated that many models
which are able to describe $(\gamma, K^+)$ experimental data tend
to unrealistically overpredict the $(\gamma, K^0)$ channel. The
use of point-like particles disregards the composite nature of
nucleons and mesons, thus losing the full complexity of a strongly
interacting hadronic system.

To provide the desired higher-energy fall-off and still preserve
the gauge invariance of the bare tree graphs, some models
introduce a cut-off function and multiply the entire
photoproduction amplitude with an overall function of monopole
form,
\begin{eqnarray}
F(\Lambda,t) ~=~ \frac{\Lambda^2-m_{K}^2}{\Lambda^2-t}
~~~~~~\textrm{or} ~~~~~~
F(\Lambda,{\bvec q}^2) ~=~ \frac{\Lambda^2-m_{K}^2}{\Lambda^2-{\bvec
q}^2}
\end{eqnarray}
where the cut-off mass $\Lambda$ is treated as a free parameter.
In spite of successfully minimizing the $\chi^2$ while maintaining
gauge invariance, there is no microscopic basis for this approach.

Field theory clearly mandates that a correct description of vertex
dressing must be done in terms of individual hadronic form factors
for each of the three kinematic situations given by the $s$-,
$u$-, and $t$-channel diagrams. In a complete implementation of a
field theory, the gauge invariance of the total amplitude is
ensured by the self-consistency of these dressing effects, by
additional interaction currents and by the effects of hadronic
scattering processes in the final state \cite{haberzettl97}.
Schematically, the interaction currents and the final-state
contributions can always be written in the form of an additional
contact diagram. If one now seeks to describe the dressing of
vertices on a more accessible, somewhat less rigorous, level, one
introduces phenomenological form factors for the individual $s$-,
$u$-, and $t$-channel vertices. Then, to ensure gauge invariance
and to remain close to the topological structure of the full
underlying theory, the simplest option is to add contact-type
currents which mock up the effects of the interaction currents and
final-state scattering processes.  In addition to the hadronic
form factors multiplying the $s$-, $u$-, and $t$-channel diagrams,
the longitudinal pieces of the gauge-invariance-preserving
additional currents are only determined up to an arbitrary
function $\tilde{F}$. For practical purposes, one of the simplest
choices \cite{haberzettl97,hbmf98} for this arbitrary function
$\tilde{F}$ has been taken to be a linear combination of the form
factors for the three kinematic situations in which the dressed
vertices appear , i.e.,
\begin{eqnarray}
{\tilde F} &=& a_sF(\Lambda ,s) + a_uF(\Lambda ,u) +
  a_tF(\Lambda ,t) ,\nonumber\\
  & &  ~~~ {\rm with~~} a_s+a_u+a_t= 1 ~,
  \label{habb_ff}
\end{eqnarray}
which introduces two more free parameters to be determined by fits
to the experimental data. This method allows fixing the $KYN$
couplings to the (approximate) SU(3) values and has proven to be
flexible and adequate for a good phenomenological description of
experimental data. It has been used in all modern studies on kaon
photoproduction in an effective Lagrangian
framework~\cite{han99,feuster99,hbmf98}. Ultimately, high-quality
data should allow an extrapolation to a Born term pole, possibly
via dispersion relations, and an extraction of the $g_{K\Lambda
N}$ and $g_{K\Sigma N}$ coupling constants.  A precise
determination of these couplings constants is especially important
in view of the fact that the $g_{\eta NN}$ coupling appears to be
much smaller than its SU(3) symmetry value \cite{tiator94}.

\subsection{Vector mesons in the $\bvec{t}$-channel}

From pion photo- and electroproduction it is well known that for
an adequate description of these electromagnetic processes the
vector meson $t$-channel contributions play an important role as
part of the background. Especially for the $p(\gamma,\pi^0)p$
reaction the $\omega$ meson is known to be an essential dynamical
ingredient. Similarly, the equivalent vector meson with
strangeness, the $K^*$(892), has been included in the description
of kaon photoproduction from the beginning. It was not until 1988,
that the pseudovector meson $K_1$(1270) was identified as also
having an important effect in model fitting \cite{adelseck88}.
Similar non-strange pseudovector mesons [the $h_1$(1170) and the
$b_1$(1235) states] have never been found to be important in pion
photoproduction. One problem with the standard vector meson
contributions in an effective Lagrangian framework is their
divergence at energies beyond $W = 2.2-2.5$ GeV. This is the
energy region which is ruled by Regge theory which starts from a
description of $t$-channel Regge trajectory exchanges at forward
angles.  These trajectories represent the exchange of a family of
mesons with the same internal quantum numbers and allow a natural
description of the smooth energy and angular dependence observed
in the data at high $W$.  Reference \cite{guidal97} has recently
applied this approach to high-energy pion and kaon photoproduction
by replacing the usual pole-like Feynman propagator of a single
particle exchanged in the $t$-channel by the so-called Regge
propagator while keeping the vertex structure given by the
effective Lagrangian for the ground state meson of the trajectory.
For the transition region between $s$-channel resonance excitation
and $t$-channel Regge exchange, duality would demand that
exchanges are limited to either all $s$-channel or all $t$-channel
contributions.  In practice, however, since no resonance model
ever includes all $s$-channel $N^*$ states the vector mesons must
be included as part of the background in some form.

\subsection{Hyperon resonances in the $\bvec{u}$-channel}

Crossing symmetry requires that the same amplitude which describes
the $p(\gamma,K^+)\Lambda$ reaction should be able to describe the
radiative capture process $p(K^-,\gamma)\Lambda$, when the
Mandelstam variables $s$ and $u$ are interchanged.  Due to SU(3)
symmetry breaking crossing symmetry in a rigorous sense cannot be
maintained since nucleons and hyperons have different excitation
spectra. Some studies~\cite{saghai96,williams92} have applied a
weaker crossing constraint by including selected hyperon
resonances in the $u$-channel of the $p(\gamma,K^+)\Lambda$
reaction and fitting the radiative capture rate,
$p(K^-,\gamma)\Lambda$, for kaons at rest along with the
$p(\gamma,K^+)\Lambda$ data.  Since hyperon resonances propagate
in the $s$-channel in radiative kaon capture this process would be
an excellent tool to constrain $Y^*$ properties. In practice, only
the capture rate at threshold has been measured; the energy
dependence of this process is unknown. On the other hand, since
hyperon resonances propagate in the $u$-channel in
$p(\gamma,K^+)\Lambda$ they contribute to the background sector -
and not to the resonance sector - of the kaon photoproduction
process. Because of the remaining uncertainties in the background
other studies \cite{mart95,feuster98,feuster99,rosenthal88} have
refrained from including hyperon resonances.

\subsection{$\bvec{S}$-channel Resonances: missing or otherwise}

One of the most contentious issues in the phenomenological
description of kaon photoproduction on the nucleon has been the
choice of s-channel nucleon resonances in the production
amplitude. Many studies have selected resonances that contribute
to the kaon photoproduction process by their relative contribution
to the overall $\chi^2$ of the
fit~\cite{saghai96,williams92,mart95,han99}. Since this is usually
done in tree-level calculations connections with other reaction
channels are difficult to establish.  As a consequence, some
studies find large couplings of the $K \Lambda$ channel to spin
5/2 resonances, even though neither recent coupled-channels
analyses \cite{feuster99,dytman2000,manley1992} nor older
partial-wave analyses for pionic $K \Lambda$
production \cite{bell83,saxon80} give any indication that such
states are important.  It is the result of these multichannel
analyses~\cite{feuster98,feuster99,dytman2000,manley1992} that
inform us of the most important resonances decaying into $K
\Lambda$ and $K \Sigma$ final states with a significant branching
ratio. In the low-energy regime the dominant resonances for the $K
\Lambda$ channel have been identified as the $S_{11}$(1650), the
$P_{11}$(1710), and the $P_{13}(1720)$ states. For the $K \Sigma$
channel, the $S_{11}$(1650) lies below threshold and the dominant
states are $p$-wave: the $P_{11}$(1710) and the $P_{13}(1720)$. At
higher energies around $W=1990$ MeV $K \Sigma$ production (both
with photons and pions) appears to be dominated by the $T=3/2$
states $S_{31}$(1900) and $P_{31}$(1910).

For the $p(\gamma, K^+)\Lambda$ channel the new {\small SAPHIR}
total cross section data \cite{saphir98} indicate for the first
time a structure around W = 1900 MeV that could not be resolved
before due to the low quality of the old data. According to the
Particle Data Book \cite{pdg98}, only the 2-star $D_{13}(2080)$
has been identified in older $p(\pi^-, K^0)\Lambda$ analyses
\cite{bell83} as having a noticeable branching ratio into the $K
\Lambda$ channel.  On the theoretical side, the constituent quark
model by Capstick and Roberts \cite{capstick94} predicts many new
states around 1900 MeV; however, only a few them have been
calculated to have a significant $K \Lambda$ decay width
\cite{capstick98} and only one, the $[D_{13}]_3$(1960), is also
predicted to have significant photocouplings \cite{capstick92}. As
discussed in more detail in Ref. \cite{mart99}, fits performed in
an isobar model lead to remarkable agreement, up to the sign,
between the quark model prediction and our extracted results for
the $D_{13}$(1960). Table \ref{tab:cc2} compares the extracted
with the predicted resonance widths not only for this "missing"(?)
state but also for the three states at lower energy. Ultimately,
only a multipole analysis will be able to unambiguously identify
the resonances contributing to kaon photoproduction. Due to the
number of double polarization observables accessible because of
the self-analyzing nature of the final hyperons a complete
experiment for this reaction may be within reach.

\begin{table}[!t]
\begin{center}
\caption{Comparison between the extracted fractional decay widths and
        the result from the quark model \protect\cite{capstick98,capstick92}
        for the $S_{11}(1650)$, $P_{11}(1710)$, $P_{13}(1720)$ and the
        "missing" $D_{13}(1900)$ resonances. }
\renewcommand{\arraystretch}{1.4}
\label{tab:cc2}
\begin{tabular}{lcc}
\hline\hline
 &\multicolumn{2}{c}{$\sqrt{\Gamma_{N^*N\gamma}
\Gamma_{N^*K\Lambda}}/\Gamma_{N^*}$ ($10^{-3}$)} \\ \cline{2-3}
Resonance~~~~~~ & ~~~~~Extracted~~~~~ & ~~~~~Quark Model~~~~~ \\ [0.5ex]
\hline $S_{11}(1650)$ & $-4.826\pm 0.051$ & $-4.264\pm 0.984$ \\
$P_{11}(1710)$ & $ ~~1.029\pm 0.172$ & $-0.535\pm 0.115$ \\
$P_{13}(1720)$ & $~ 1.165^{+0.041}_{-0.039} $ & $-1.291\pm
0.240$\\ $D_{13}(1900)$ & $~ 2.292^{+0.722}_{-0.204} $ &
$-2.722\pm 0.729$\\[0.5ex]
\hline\hline
\end{tabular}
\end{center}
\end{table}

\subsection{An effective tree-level operator for nuclear applications}

While it is generally recognized that a detailed understanding of
the various reactions participating in resonance production must
take place within a multichannel framework, these amplitudes are
much too cumbersome to use in the nuclear environment.  Nuclear
calculations of pion, eta or kaon photoproduction usually need
amplitudes that are easy to incorporate, have a straightforward
off-shell extension, can be boosted to different Lorentz frames
and are fast in terms of computer time. This is especially true
for few-body calculations where high-dimensional integrals are
solved in momentum space, requiring the subroutine for the
elementary operator to be used on the order of $10^6$ times. For
this purpose the elementary amplitude is then parameterized in a
form that makes it easy to use, describes the elementary data well
and preserves as many of the field-theoretical constraints as
possible.  In general, this involves constructing a tree-level
amplitude, selecting a limited number of resonances which have
been shown to be significant by multichannel analyses and
refitting the coupling constants to the experimental data. Work
over the last several years has shown that restoring gauge
invariance in the presence of form factors is straightforward at
the tree level. However, neglecting the final meson-baryon
interaction in the meson photoproduction amplitude automatically
leads to a violation of unitarity which is more difficult to
restore. One possibility is to allow for energy-dependent phases
for each multipole, as is done in the Mainz MAID pion
electroproduction isobar model \cite{kamalov1999}. Furthermore,
imposing dispersion relations would help to establish the proper
analyticity properties of such effective amplitudes.  No such
attempts have yet been done for kaon photo- and electroproduction.

\section{\bf \normalsize HYPERNUCLEAR EXCITATION}

\begin{figure}
\begin{minipage}[htb]{75mm}
\centerline{\psfig{file=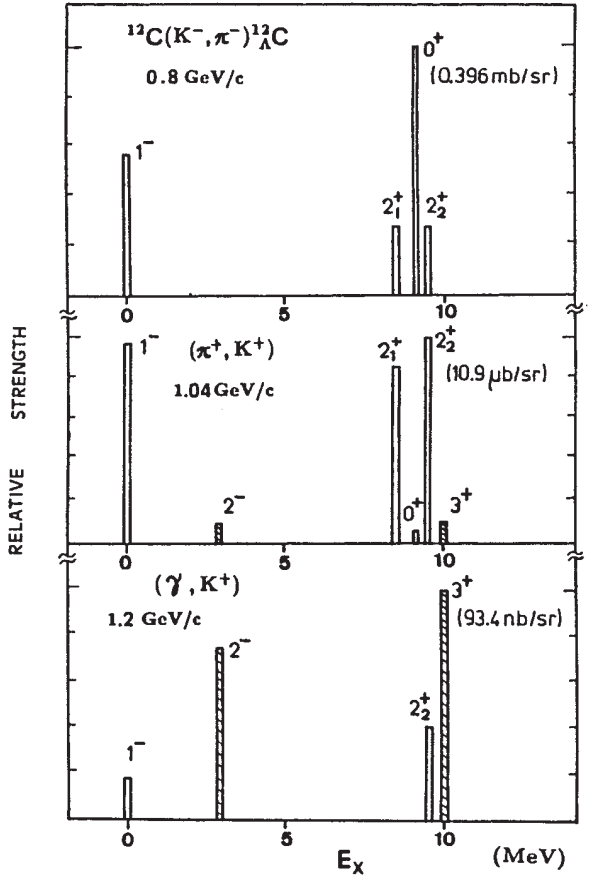,width=8cm}}
\end{minipage}
\hspace{\fill}
\begin{minipage}[htb]{75mm}
\caption{Comparison
of the theoretical excitation spectra for the reactions
$(K^-,\pi^-)$, $(\pi^+,K^+)$ and $(\gamma,K^+)$, producing
$^{12}_{~\Lambda}$C ($^{12}_{~\Lambda}$B) at $\Theta_K =
10^\circ$. The $1^-$ and $2^-$ states belong to the ground state
configuration $(p_{3/2}^{-1},_{\Lambda}s_{1/2})1^-, 2^-$ while the
levels around $E_x$=10 MeV correspond to a $\Lambda$ in a
$p_{3/2}$ or $p_{1/2}$ orbit with levels
$(p_{3/2}^{-1},_{\Lambda}p_{3/2})0^+,1^+_1,2^+_1,3^+$ and
$(p_{3/2}^{-1},_{\Lambda}p_{1/2})1^+_2,2^+_2$. Only the strongly
excited states are shown. This figure is taken from
Ref.\,\protect\cite{bando90}.} \label{hyplabf1}
\end{minipage}
\end{figure}

With the recent successful completion of the Jlab Hall C
experiment 89-009 \cite{exp89-009} which produced discrete
hypernuclear states with electrons for the first time, the
exploration of hypernuclear structure through electromagnetic
probes is becoming a reality. In contrast to the hadronic
reactions $(K^-, \pi^-)$ and $(\pi^+,K^+)$, the $(\gamma, K^+)$
process uses, besides the photon, the rather weakly interacting
$K^+$ with its mean free path of 5-7 fm in the nuclear medium,
allowing the process to occur deep in the nuclear interior.  In
comparison, the $K^-$ and the $\pi^{\pm}$ are both strongly
absorbed, thereby confining the reaction to the nuclear periphery.
Due to the mass difference in the incoming kaon and outgoing pion,
the $(K^-,\pi^-)$ reaction allows for recoilless $\Lambda$
production in the nucleus, leading to high counting rates.  Kaon
photoproduction, on the other hand, involves high momentum
transfers due to the large production of the rest mass which will
therefore project out high momentum components of the nuclear wave
functions.  Figure~\ref{hyplabf1} illustrates the differences
between the $(K^-, \pi^-)$, $(\pi^+,K^+)$, and $(\gamma, K^+)$
production reactions, both in terms of their relative excitation
strength and in terms of the structure of the produced
hypernuclear spectrum for p-shell hypernuclei. At a strength of
several 100 mb/sr the $(K^-, \pi^-)$ reaction predominantly
excites natural parity states with low angular momentum, such as
the ground state $(p_{3/2}^{-1},_{\Lambda}s_{1/2})1^-$ or the
substitutional state $(p_{3/2}^{-1},_{\Lambda}p_{3/2})0^+$.
Reduced by about a factor of 50 in cross section, the
$(\pi^+,K^+)$ reaction still excites natural parity levels, but
selecting the ones with large angular momentum, such as the
$(p_{3/2}^{-1},_{\Lambda}p_{3/2})2^+_1$ and
$(p_{3/2}^{-1},_{\Lambda}p_{1/2})2^+_2$ states, reflecting the
larger momentum transfer of the process.  Finally, the $(\gamma,
K^+)$ reaction excites primarily the unnatural parity, high
angular momentum states, such as the ground state
$(p_{3/2}^{-1},_{\Lambda}s_{1/2})2^-$ or the substitutional state
$(p_{3/2}^{-1},_{\Lambda}p_{3/2})3^+$, albeit with a strength
reduced by another two orders of magnitude, as one would expect
for the electromagnetic interaction.  This comparison demonstrates
that full spectroscopic information can only be obtained with a
combination of all three techniques. The subject of exciting
discrete hypernuclear states through kaon photoproduction was
studied extensively about 8-10 years ago
\cite{bennhold89,rosenthal88,cotanch86,cohen89} but has been
mostly dormant for the last several years, awaiting data taking.
Therefore, the number of planned and approved experiments to take
place within the next few years is expected to revive interest in
this field.

\subsection{Matrix elements for the process $\gamma + {\rm A}
\rightarrow K^+ + ~_\Lambda{\rm B}$ }

As shown in detail in Ref.~\cite{bennhold89}, assuming a one-body
kaon photoproduction operator, the many-body nuclear matrix
element naturally separates into a nuclear structure piece and a
single-particle piece:

\begin{eqnarray}
\langle J_fM_f;K^+|T|J_iM_i;\gamma \rangle &=& \sum_{\alpha ,
\alpha '} \langle J_fM_f;K^+|C_{\alpha '}^\dagger C_{\alpha}
|J_iM_i;\gamma \rangle~  ~\langle \alpha '
;K^+|t|\alpha ;\gamma \rangle ~. \label{eq:basic}
\end{eqnarray}
In Eq.~(\ref{eq:basic}) the many-body nuclear structure aspects
are separated from the photoproduction mechanism but in principle
the sum extends over a complete set of single-particle states
$\alpha$ and $\alpha '$. The nuclear structure information
involved in one-body processes is usually contained in the reduced
density matrix elements (RDME),
\begin{eqnarray}
\Psi_{J}(a',a) &=& {\hat J}^{-1} \langle J_f || [C_{\alpha
'}^\dagger \otimes C_{\alpha}] ||J_i \rangle ~.
\end {eqnarray}

All the dynamics of the photoproduction process is contained in the
single-particle matrix element $\langle \alpha ' ;K^+|t|\alpha ;\gamma
\rangle $ which in general involves a nonlocal
operator. In momentum space this matrix element has the form
\begin{eqnarray}
\langle \alpha ' ;K^+|~t~|\alpha ;\gamma \rangle &=& \int d\!~^3p~ d\!~^3q'~
{\psi}^{*}_{\alpha '}(\bvec{p}') \phi^{*(-)}_{K}
(\bvec{q},\bvec{q}')~ t_\gamma~ \psi_{\alpha}(\bvec{p}) ~,
\label{sixdim}
\end{eqnarray}
where $\bvec{p}' = \bvec{p} + \bvec{k} - \bvec{q}$, and $\psi$ is the single-particle
wave function of the proton in the initial and the $\Lambda$ in the final state.
The wave function with the
appropriate boundary conditions for the outgoing kaon of three-momentum
$\bvec{q}$, distorted by its interaction with the residual hypernucleus
through an optical potential, is denoted by $\phi^{*(-)}_K (\bvec{q},\bvec{q}')$.
This wave function is generated by solving the Klein-Gordon
equation using a simple $t \rho$ optical potential with the $K^+ N$ phase shifts
of Ref.~\cite{martin75}.

\subsection{Basic features of the coherent kaon production
process}

\begin{figure}
\begin{minipage}[htb]{75mm}
\centerline{\psfig{file=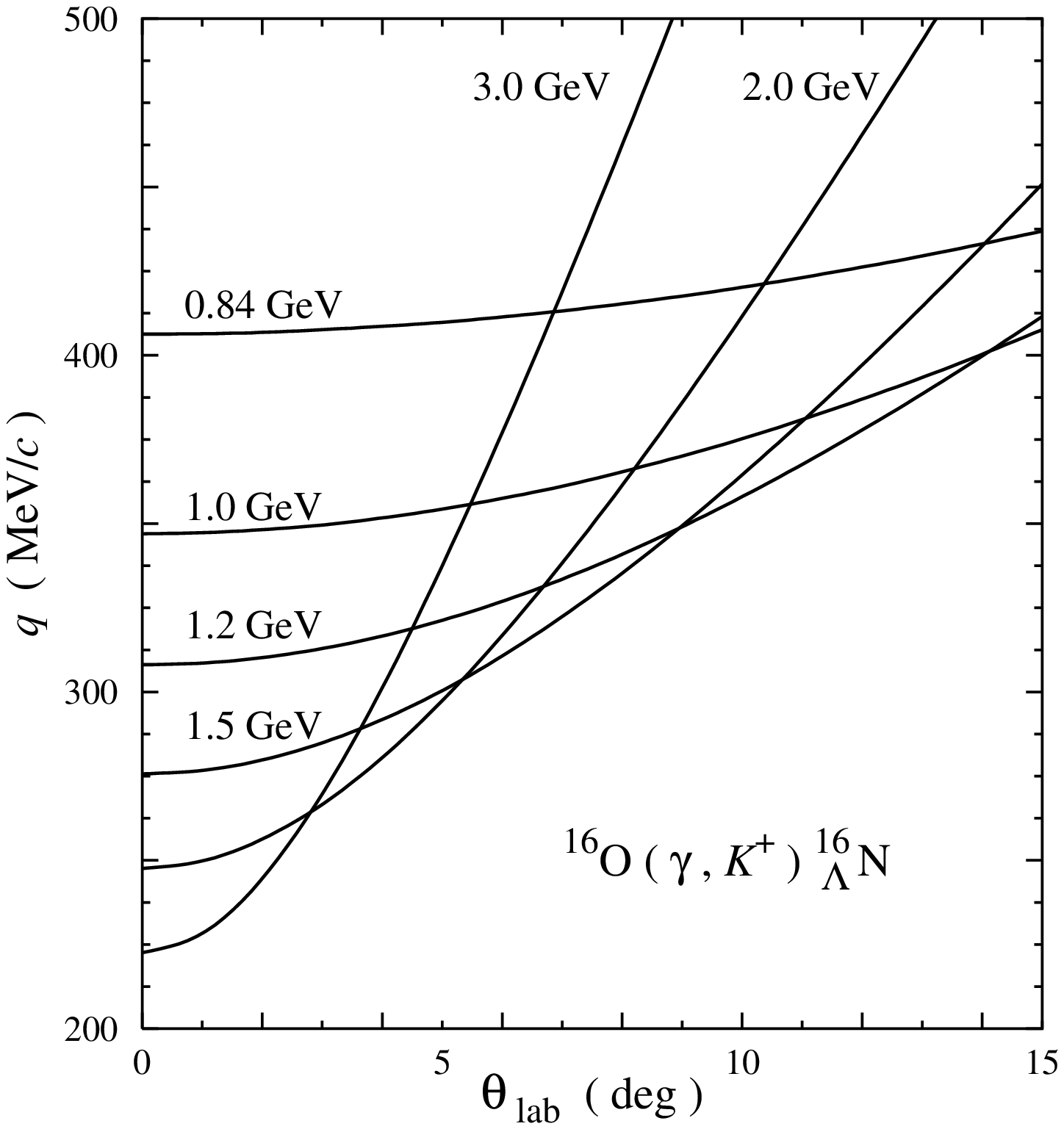,width=77mm}}
\end{minipage}
\hspace{\fill}
\begin{minipage}[htb]{75mm}
\centerline{\psfig{file=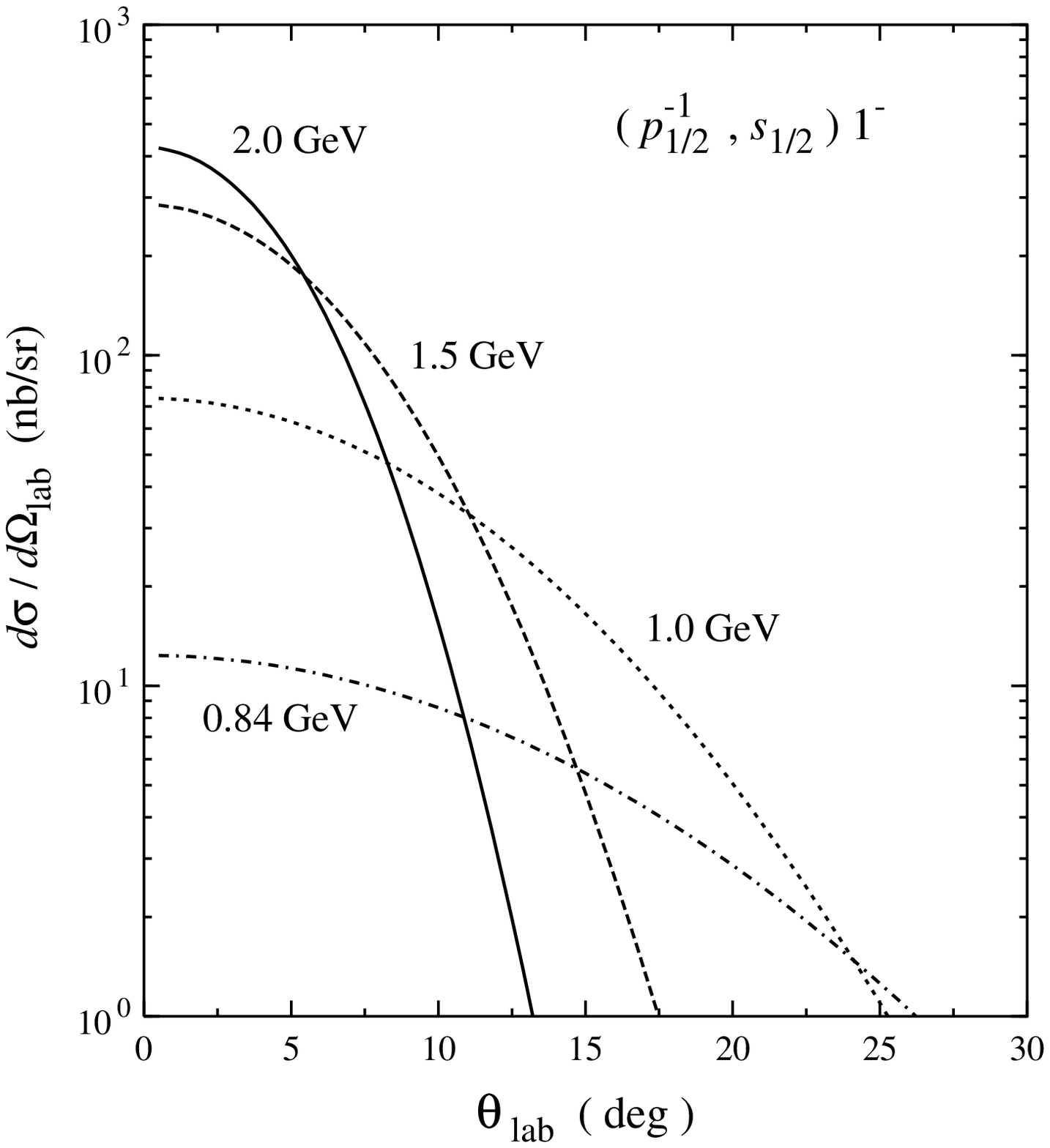,width=77mm}}
\end{minipage}
\caption{Kinematic features of the $(\gamma, K^+)$ process. The
left side shows the momentum transfer in the lab system for
several photon energies $E_{\gamma}$.  The right side shows
angular distributions for the $1^-$ member of the ground state
doublet in $^{16}{\rm O}(\gamma,K^+)^{16}_{~\Lambda}{\rm N}$ for
various photon lab energies.} \label{hyplabf2}
\end{figure}

Figure \ref{hyplabf2} compares the momentum transfer behavior with
the magnitude of the differential cross section for the reaction
$^{16}{\rm O}(\gamma,K^+)^{16}_{~\Lambda}{\rm N}$. At $0^\circ$
kaon lab angle the momentum transfer to the final hypernuclear
system decreases as the photon lab energy increases.  This leads
to a differential cross section at $\Theta_K=0^\circ$ which
increases as $E_{\gamma}$ increases, from around 15 nb/sr at 0.84
GeV to 330 nb/sr at 2 GeV for the particular transition shown.
However, the momentum transfer increases more rapidly for non-zero
kaon angles at higher photon energies. Thus, the angular
distributions become more forward peaked and fall off more
rapidly.  The energy chosen for an experiment therefore depends on
the desired result: If the goal is to perform hypernuclear
spectroscopy choosing a higher photon energy around 2 GeV, while
detecting the $K^+$ under $0^\circ$ would be advantageous. If, on
the other hand, one likes to extract dynamical information by
mapping our transition densities via measuring angular
distributions, photon energies of 1.0 to 1.2 GeV are preferable.

\subsection{Extracting hypernuclear structure information}

The hypernuclear production experiments over the last decades have
found hypernuclear ground and excited states that can be
reproduced well within the weak coupling model which assumes that
the $\Lambda$ couples weakly to the ground and excited states of
the core nucleus.  Studying these states through hypernuclear
spectroscopy reveals details of the effective $YN$ interaction in
nuclear matter.  Using appropriate nuclear $G$-matrix techniques
in principle allows a self-consistent extraction of the elementary
two-body $YN$ interaction.

\begin{figure}
\centerline{\psfig{file=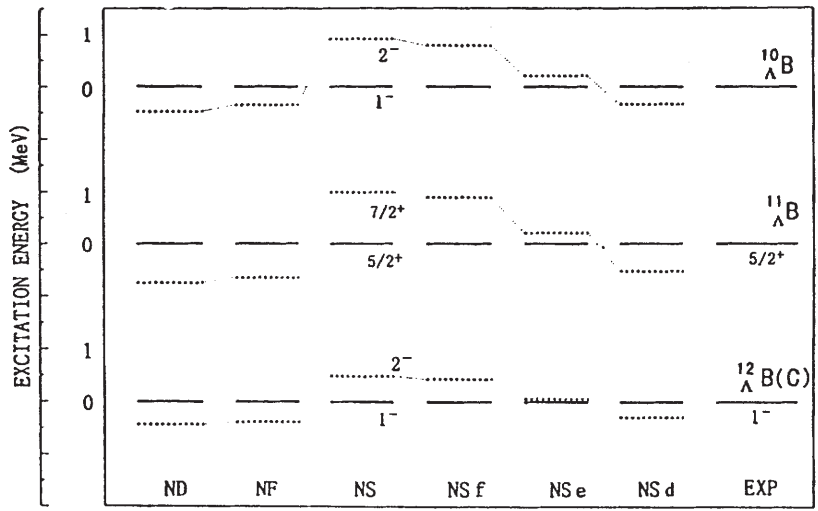,width=11cm}}
\caption{Spin-doublett states in $^{10}_{~\Lambda}B$,
$^{11}_{~\Lambda}$B, and $^{12}_{~\Lambda}$B (C) for the Nijmegen
potentials D (ND), F (NF), NSC89 (NS), and NSC97d, e, and f,
respectively. This figure is taken from Ref.~\cite{yamamoto}.}
\label{hyplabf3}
\end{figure}

Figure \ref{hyplabf3} demonstrates the power of such an approach:
Comparing six different Nijmegen $YN$ potentials in hypernuclear
$G$-matrix calculations Ref.~\cite{yamamoto} finds that for the
ground state splittings of $^{11}_{~\Lambda}$B and
$^{12}_{~\Lambda}$B only the NSC89, and the NSC97e and f
interaction have spin-spin interactions repulsive enough to
reproduce the correct ordering of the states.  This finding is
supported by few-body calculations where it is found that only the
above interactions can bind the hypertriton.  Experimental
resolution of the ground state splittings shown in
Fig.~\ref{hyplabf3} would allow discriminating further between
those three forces. The position of higher-lying excited states is
sensitive to the $\Lambda N$ spin-orbit and tensor interaction.

With the exception of Refs.~\cite{rosenthal88,motoba94} all
calculations up to now have been performed in pure particle-hole
configurations. These predictions may be reliable  where the
proton pick-up strength is not highly fragmented for stretched
spin-flip transitions with maximum $J=l_N + l_{\Lambda} +1$. These
transitions are usually dominated by the Kroll-Ruderman {\boldmath
$\sigma\cdot\epsilon$} operator and tend to have the largest cross
sections. In the case shown in Fig.~\ref{hyplabf2}
$^{16}_{~\Lambda}$N is described as a pure $p_{1/2}$ proton hole
coupled to an s-shell $\Lambda$, coupling to a $0^-$ and $1^-$
ground state transition.  For a closed shell target nucleus in a
pure particle-hole basis, the RDME simply reduce to
$\Psi_{J;T}(a',a) = \delta_{ab} \delta_{a'b'}$. The degeneracy
between the two ground states would be removed by including the
residual $\Lambda N$ interaction. For $p$-shell hypernuclei with
the $\Lambda$ in an $s$ orbit, the $p_N s_\Lambda$ interaction can
be expressed in terms of the five radial integrals $\bar V$,
$\Delta$, $S_\Lambda$, $S_N$, and $T$, assumed to be constant
across the $p$-shell and associated with the average central,
spin-spin, $\Lambda$ spin-orbit, induced nucleon spin-orbit and
tensor terms in the potential \cite{millener85}:

\begin{eqnarray}
V_{\Lambda N} (r) &=& V_0(r) + V_\sigma (r)\, \bvec{s}_N\cdot
\bvec{s}_\Lambda + V_\Lambda(r)\,
\bvec{l}_{N\Lambda}\cdot\bvec{s}_\Lambda + V_N(r) \,
\bvec{l}_{N\Lambda}\cdot\bvec{s}_N + V_T(r)\, S_{12} ~.
\end{eqnarray}

Performing an analysis of hypernuclear structure data the
``standard interaction'' of Ref. \cite{millener85} uses the
following values (in MeV): $\Delta=0.50$, $S_\Lambda =-0.04$,
$S_N=-0.08$, and $T=0.04$. Doublet splittings are determined
mainly by the spin-spin, $\Lambda$ spin-orbit, and tensor
interactions $\Delta$, $S_\Lambda$, and $T$, leading to the
following expressions for the separation energies of the
particle-hole pairs:

\begin{eqnarray}
p_{3/2}s_{1/2} ~~~~~~ \delta ~=~ {\textstyle
\frac{2}{3}}\,\Delta + {\textstyle \frac{4}{3}}\, S_{\Lambda} -
{\textstyle \frac{8}{5}}\, T ~~.
\end{eqnarray}
and
\begin{eqnarray}
p_{1/2}s_{1/2} ~~~~~~ \delta ' ~=~ -{\textstyle
\frac{1}{3}}\,\Delta + {\textstyle \frac{4}{3}}\, S_{\Lambda} + 8T
~~.
\end{eqnarray}

\begin{figure}
\centerline{\psfig{file=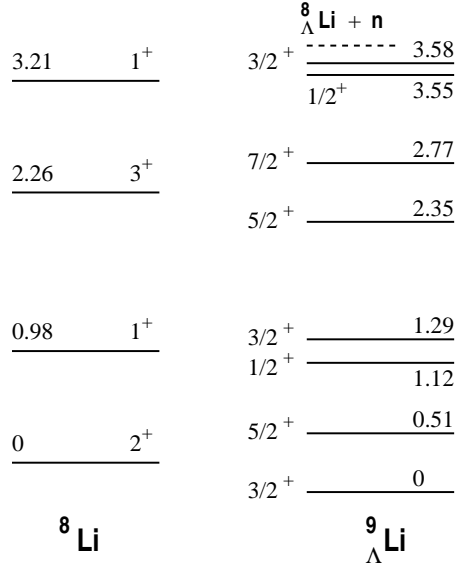,width=6cm}} \caption{The energy
levels (in MeV) for $^8$Li from experiment and for
$^{9}_{\Lambda}$Li calculated from the standard interaction by
Millener \cite{millener85}. The isospin partner $^{9}_{\Lambda}$Be
could be reached using the reaction $^{9}{\rm
Li}(\gamma,K^+)^{9}_{\Lambda}{\rm Be}$. This figure is taken from
Ref.~\cite{millener90}.} \label{hyplabf4}
\end{figure}

This results in $\delta = 216$ keV and $\delta '=100$ keV; thus
doublet splittings are generally small.
While the stretched transitions are most likely the first ones to
be measured, eventually one would like to use the $(\gamma, K^+)$
reaction to extract hypernuclear structure information from cases
where configuration mixing is important. As discussed in
Ref.~\cite{millener90}, the reaction $^{9}{\rm
Be}(\gamma,K^+)^{9}_{~\Lambda}{\rm Li}$ may provide a good testing
ground for resolving members of the $s_{\Lambda}$ doublet of
$3/2^+$ and $5/2^+$, coupling the s-shell $\Lambda$ to the $2^+$
core of ${^8}$Li. As shown in Fig.~\ref{hyplabf4}, this doublet is
split by 0.51 MeV according to the "standard" $\Lambda N$
interaction of Ref.~\cite{millener85}.  This splitting is mainly
due to the the spin-spin part of the $\Lambda N$ interaction,
thus, resolving these states would provide an improved constraint.
The predicted $\Delta S$ = 0 RDME are large for the lower member
of the doublet but small for the upper member while there is a
large $\Delta J$ = 2 ($\Delta S$ = 1) RDME for the upper member;
the two transitions would therefore produce very different angular
distributions. Similar information may be extracted from the
reaction $^{13}{\rm C}(\gamma,K^+)^{13}_{~\Lambda}{\rm B}$.

\begin{figure}
\centerline{\psfig{file=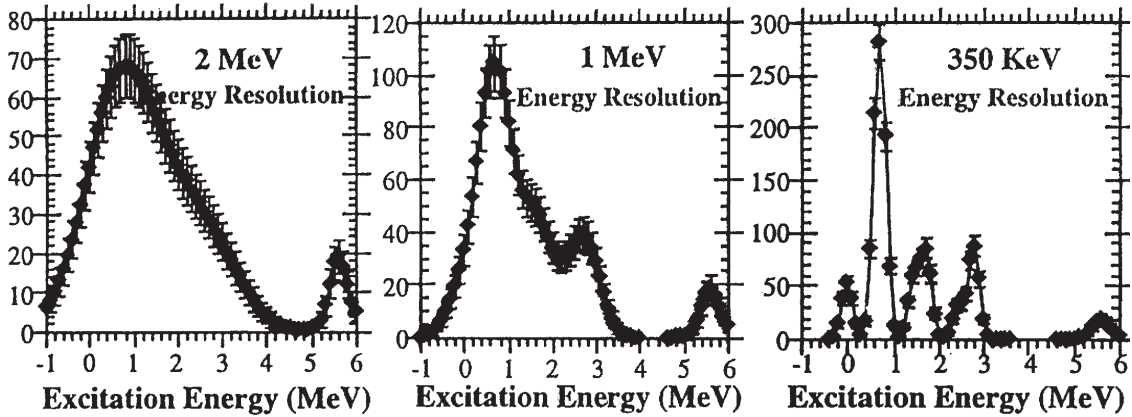,width=15cm}} \caption{Simulated
excitation function for the reaction $^{9}{\rm
Li}(e,e'K^+)^{9}_{\Lambda}{\rm Be}$ with increasing experimental
resolution of 2 MeV, 1 MeV, and 0.35 MeV. This figure is taken
from Ref.~\cite{jlab-e94-107}.} \label{hyplabf5}
\end{figure}

Overall, many energy splittings of doublets with Lambdas in the
s-orbit are predicted to lie well below 100 keV and may only be
resolvable through additional $\gamma$-ray spectroscopy.  However,
a few doublet splittings, due mainly to the spin-spin force, are
predicted to be around 500 keV and should be observable with kaon
detectors planned for Hall A. Peaks due to $s_{\Lambda}$ doublets
based on different nuclear core states should be well separated.
Furthermore, coupling $p_{\Lambda}$ orbitals to nuclear core
states produces a large number of excited states, predicted to be
separated by several hundred keV.  High resolution spectrometers
for the kaons are therefore imperative in order to make progress
in this field. Figure~\ref{hyplabf5} demonstrates in a simulation
of a planned Jlab Hall A detector the appearance of closely spaced
levels with increasing experimental resolution.

\subsection{Maping out the $\Lambda$ wave function in the
hypernucleus}

Stretched transitions of the sort
$(p_{3/2}^{-1},_{\Lambda}s_{1/2})2^-$ or
$(p_{3/2}^{-1},_{\Lambda}p_{3/2})3^+$ can be predicted almost
model independently since they are dominated by the {\boldmath
$\sigma\cdot\epsilon$} Kroll-Ruderman term. If a transition is
dominated by the Kroll-Ruderman term, the operator becomes local
and can be factored out of the single-particle matrix element of
Eq.~(\ref{sixdim}). Furthermore, neglecting kaon distortion which
reduces cross sections only by about 10-20\% for $p$-shell nuclei
reduces the matrix element to
\begin{eqnarray}
\langle \alpha ' ;K^+|t|\alpha ;\gamma \rangle & = &
\textrm{const.} \int r^2 dr~
{\psi}^{*}_{\Lambda}(r)  \psi_{p}(r) j_{L}(Qr) ~.
\end{eqnarray}

\begin{figure}
\centerline{\psfig{file=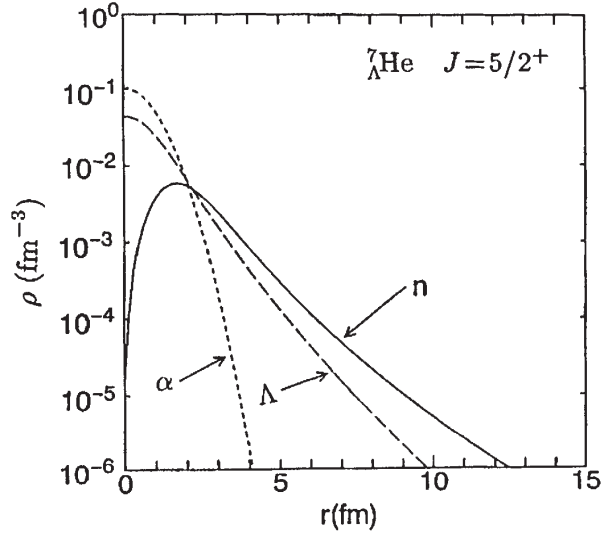,width=8cm}} \caption{Density
distributions of the valence nucleon and $\Lambda$ in the weakly
bound state of $^{7}_{\Lambda}{\rm He} (5/2^+)$.  The radius $r$ is
measured from the $^{5}_{\Lambda}{\rm He}$ cm frame. This figure
is taken from Ref.~\cite{hiyama96}.} \label{hyplabf6}
\end{figure}

Therefore, assuming one has good knowledge of the bound proton
wave function from $(e,e' p)$ experiments, measuring a kaon
angular distribution will be sensitive to the bound $\Lambda$ wave
function. Such information may prove especially valuable in
certain few-body cases where adding a $\Lambda$ "impurity" to the
nucleus can lead to significant rearrangement of the nucleus.  As
discussed in detail in Ref.~\cite{hiyama96} for the A=6 and 7
hypernuclei, this new dynamical feature can lead to new bound
states and appreciable contraction of the entire nuclear system.
For example, if a $\Lambda$ is added to the weakly-bound halo
nucleus $^6$He, the predicted core-neutron rms radius decreases
from 4.55 fm to 3.55 fm.  The ground-state binding energies move
from -0.98 MeV for $^6$He (measured from the  $^4{\rm He} + 2n$
threshold) to -2.83 MeV for $^7_{\Lambda}$He  (predicted with
respect to $^6_{\Lambda}{\rm He} +n$ breakup), changing the
nucleus from a Borromean to a non-Borromean system.
Fig.\,\ref{hyplabf6} shows the density distributions of the
$\alpha$-core, the $\Lambda$ skin and the neutron halo within
$^7_{\Lambda}$He.  Using the reaction $^{7}{\rm
Li}(\gamma,K^+)^{7}_{\Lambda}{\rm He}$ one could access this
exotic hypernucleus and map out the bound state $\Lambda$ wave
function. This feature is unique to the $(\gamma, K^+)$ process
since distortion effects are minimal.

\subsection{$\Sigma$ hypernuclei}

\begin{figure}
\centerline{\psfig{file=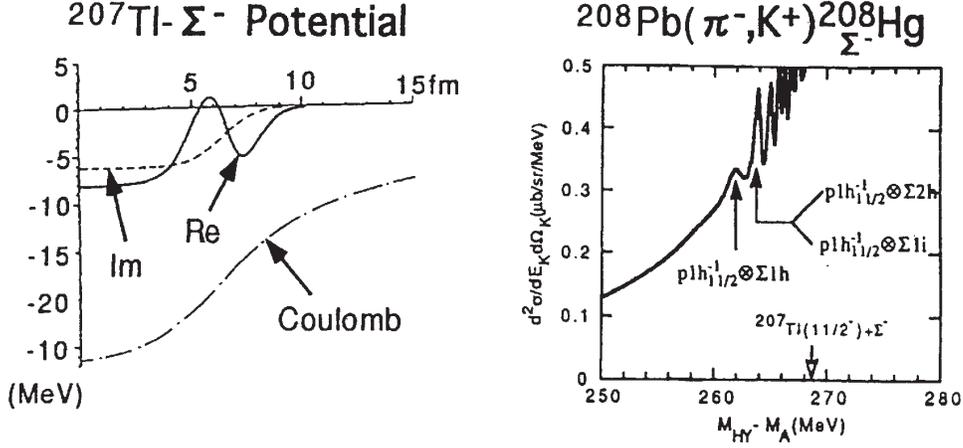,width=13cm}} \caption{Calculated
$\Sigma$-Tl potential (left) and the $^{208}_{\Sigma^-}$Hg
spectrum produced with the $(\pi^-,K^+)$ reaction. This figure is
taken from Ref.~\cite{tadokoro92}.} \label{hyplabf7}
\end{figure}

Little is known about the $\Sigma N$ interaction and the
$\Sigma$-nucleus potential.  Old bubble-chamber analyses revealed
a magnitude comparable to the $\Lambda N$ interaction, with a
significant role played by strong $\Sigma N \rightarrow \Lambda N$
conversion.  First quantitative results on the $\Sigma$-nucleus
potential were obtained by $\Sigma^-$ atom x-ray data, yielding a
potential in the nuclear center of $-$(25-30) MeV for the real
part and $-$(10-15) MeV for the imaginary part.  If such an
analysis is performed with a potential nonlinear in the nuclear
density the resulting real part of the $\Sigma$-nucleus potential
becomes very shallow or even repulsive \cite{batty94}. In this
context, the formation of bound states was considered unlikely.
However, to the surprise of the community, in 1980 the
Saclay-Heidelberg group reported narrow structures in the unbound
region of the $^9{\rm Be}(K^-,\pi^-)$ spectrum \cite{bertini80}.
Follow-up experiments at KEK and BNL were unable to verify this
finding, except for the $A=4$ system.  Using the $^4{\rm
He}(K^-,\pi^-)$ reaction a clear signal for a  $^4_{\Sigma}$He
bound state was observed \cite{hayano89,nagae98} with a binding
energy of $E_x = 4.4$ MeV and a width of $\Gamma = 7.0$ MeV.  The
unique nature of this state appears to be due to the strong
isospin dependence of the $\Sigma$-nucleus potential, with the
$T=1/2$ part consisting of a repulsive core and an attractive
pocket near the nuclear surface.  The presence of the repulsive
core reduces the wave function overlap between the $\Sigma$ and
the residual nucleus, leading to a suppression of the $\Sigma N
\rightarrow \Lambda N$ conversion width. Kaon photoproduction
would allow investigating the $A=4$ hypernuclear system via the
process $^{4}{\rm He}(\gamma,K^+)^{4}_{\Sigma}{\rm H}$. In heavy
systems the isospin-independent part is expected to dominate the
$\Sigma$-nucleus potential.  Shown in Fig.\,\ref{hyplabf7} is a
calculation by Ref.~\cite{tadokoro92}, predicting an attractive
potential in the nuclear interior with a repulsive bump near the
surface.  Combing such a strong potential with the Coulomb
potential can produce narrow structures, as shown in the right
panel of Fig.\,\ref{hyplabf7}.  While the predictions of
Ref.~\cite{tadokoro92} shown in Fig.\,\ref{hyplabf7} were
obtained using the $(\pi^-, K^+)$ reaction, we propose to rather
photoproduce such Coulomb-assisted $\Sigma$-hypernuclear bound
states through reactions such as $^{208}{\rm
Pb}(\gamma,K^+)^{208}_{\Sigma^-}{\rm Tl} $.  This would eliminate
the need for a strongly absorbed pion in the initial state and
open the possibility to populate deeply bound
$\Sigma$-hypernuclear states.  Again, high-resolution detectors
are imperative for this kind of second-generation experiment.

\subsection{A special case: Photoproduction of the hypertriton}

In the periodic table of hypernuclei, the hypertriton, a bound
state consisting of a proton, a neutron, and a lambda, holds a
special place as the lightest hypernucleus. Since neither the
$\Lambda N$ nor the $\Sigma N$ interactions are sufficiently
strong to produce a bound two-body system with $S=-1$, the
hypertriton is the first system in which the $YN$ force, including
the interesting $\Lambda$-$\Sigma$ conversion potential, can be
tested in the nuclear environment. Therefore, the hypertriton
plays an important role in hypernuclear physics, similar to the
deuteron in nuclear physics. Ref.~\cite{miyagawa2} has carried
out a detailed investigation of this system using Faddeev
equations and found that binding the hypertriton with its small
binding of 130 keV (with respect to $\Lambda - d$ break-up),
requires a fine-tuning of the $YN$ $^1S_0$ amplitude which is
found in only few potentials, such as the NSC89\cite{nijmegen1}.
While this feature makes the hypertriton a fascinating system to
study, it is precisely the weak binding that makes this loosely
bound system, which displays halo-like features similar to
neutron-rich nuclei near the neutron drip line, a difficult system
to produce in high-momentum reactions such as $^3{\rm
He}(\gamma,K^+)^3_{\Lambda}{\rm H}$.

In contrast to the transition matrix elements for $p$-shell
hypernuclei [Eq.\,(\ref{eq:basic})], the nuclear matrix elements for the
process on $^3$He do not separate into a many-body nuclear
structure part and a single-particle one-body integral, but they
are evaluated straightforwardly in terms of an integral over all
internal momenta and states contributing to the process
\cite{mart98},
\begin{eqnarray}
T_{\rm fi} &=& \langle~ ^3_{\Lambda}{\rm H}~|~ t^{\gamma p \rightarrow
K^+  \Lambda} ~|~ ^3{\rm He}~ \rangle ,\nonumber\\
& = & \left( \frac{E_{\rm ^{3}He} E_{\rm ^{3}_{\Lambda}H}}{M_{\rm
^{3}He} M_{\rm ^{3}_{\Lambda}H}} \right)^{1/2} \int d^{3}{\bvec p}~d^{3}
{\bvec q}
\left(\frac{m_{\rm f} m_{\rm i}}{E_{\rm f} E_{\rm i}} \right)^{1/2}
\Psi_{\rm ^{3}_{\Lambda}H}({\bvec p}, {\bvec q}\, ')~
t^{\gamma p \rightarrow K\Lambda}({\bvec q}, {\bvec Q})~
\Psi_{\rm ^{3}He}({\bvec p}, {\bvec q}\, ),
\label{tfi33}
\end{eqnarray}

In order to estimate the magnitude of the production cross section
one may consider the struck nucleon inside $^3$He
as having a fixed momentum \cite{kamalov}. In this case,
the elementary operator
can be factored out of the integral and the cross
section off $^3$He may be written as
\begin{eqnarray}
  \label{msabit}
 \frac{d\sigma_{\rm T}}{d\Omega_K} &=&{\textstyle \frac{1}{9}}~ W_A^2~|F(Q)|^2
~\left( \frac{d\sigma_{\rm T}}{d\Omega_K}\right)_{\rm proton} ~,
\end{eqnarray}
with the nuclear form factor
\begin{eqnarray}
  \label{efq1}
F(Q) &=& \int~d^3{\bvec q}~d^3{\bvec p}~~
\Psi_{^3_{\Lambda}{\rm \!H}}({\bvec p},{\bvec q}+{\textstyle \frac{2}{3}}
{\bvec Q})~~ \Psi_{^3{\rm He}}({\bvec p},{\bvec q}\, ) ~,
\end{eqnarray}

\begin{figure}
\begin{minipage}[htb]{75mm}
\centerline{\psfig{file=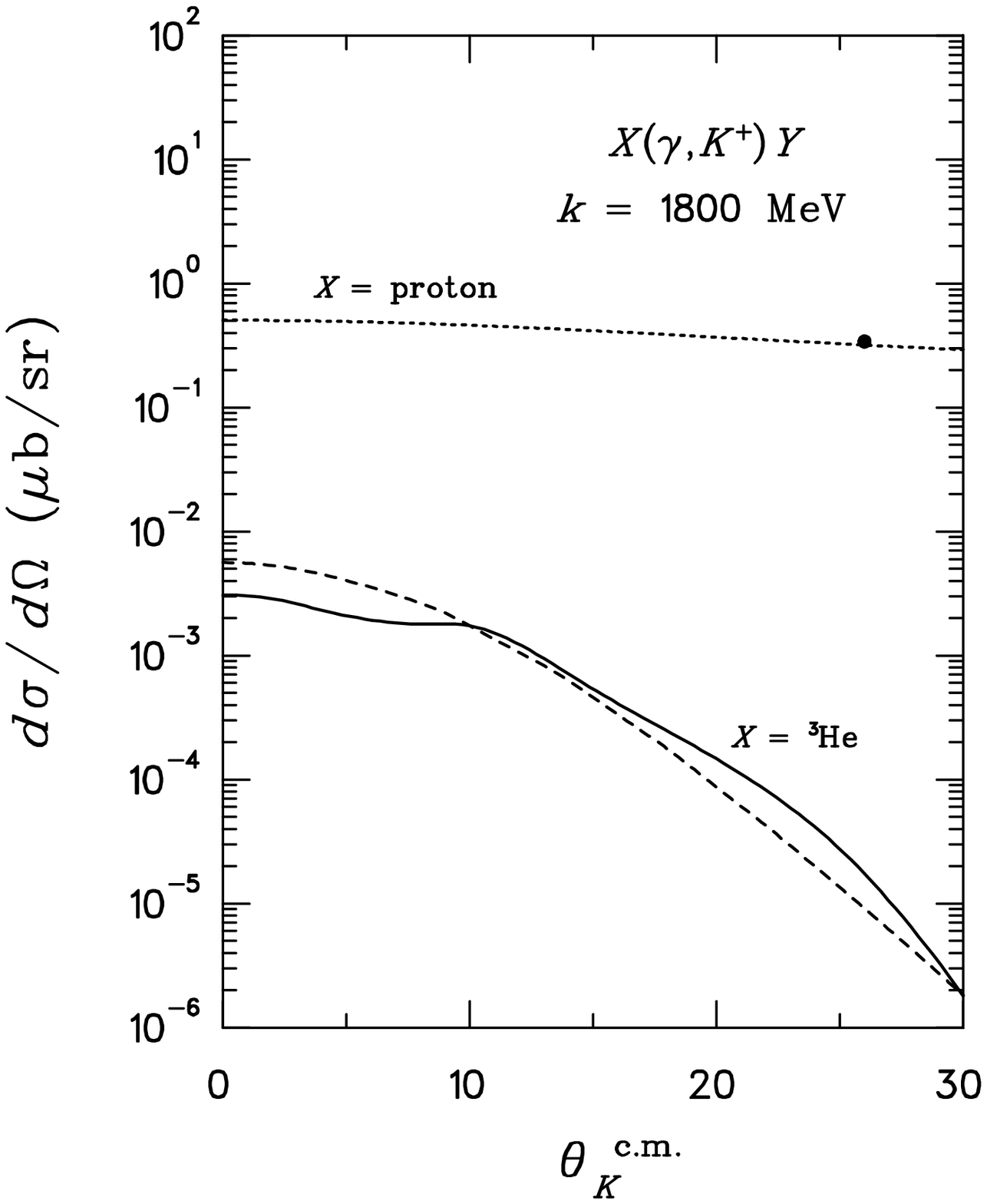,width=75mm}}
\caption{Differential cross section for kaon photoproduction off
the proton
 and $^3$He as function of kaon angle. The elementary reaction (dotted
 line) is taken from  Ref.~\protect\cite{williams92} and the corresponding
 experimental datum is from Ref.~\protect\cite{fe}. The dashed line shows the
 approximation for production off $^3$He calculated from Eq.~(\ref{msabit}),
 the solid line represents the exact calculation using $S$-waves.
 The figure is taken from Ref.~\cite{mart98}.}
\label{hyplabf8}
\end{minipage}
\hspace{\fill}
\begin{minipage}[htb]{75mm}
\centerline{\psfig{file=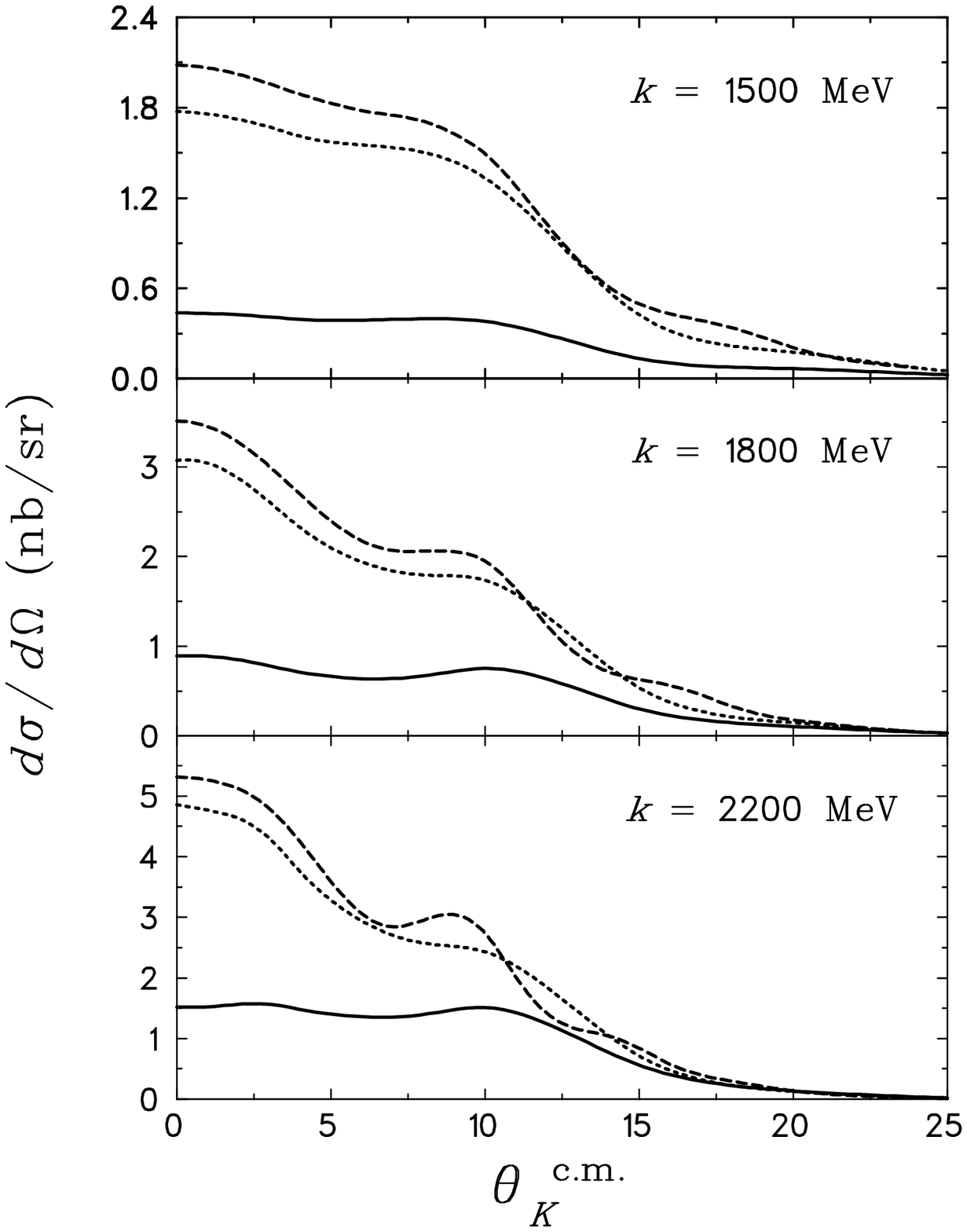,width=70mm}}
\caption{The cross section for kaon photoproduction off $^3$He at three
 different excitation energies. The dotted curves are obtained from the
 the calculation with $S$--waves only and the simple hypertriton wave
function, the dashed curves are obtained with $S$--waves only and
the correlated Faddeev wave function of
Ref.~\protect\cite{miyagawa2}, while the solid curves show the
result after using all of the partial waves and the simple
hypertriton wave function\cite{congleton}.
 The figure is taken from Ref.~\cite{mart98}.}
\label{hyplabf9}
\end{minipage}
\end{figure}

As shown in Fig.\,\ref{hyplabf8}, using the nuclear form factor of
Eq.~(\ref{efq1}) reduces the reaction cross section of
Eq.~(\ref{msabit}) by two orders of magnitude compared to the
elementary reaction. As $\theta_K^{\rm c.m.}$ increases, the cross
section drops quickly, since the nuclear momentum transfer
increases rapidly as function of $\theta_K^{\rm c.m.}$ (see
Fig.~\ref{hyplabf2}). Figure \ref{hyplabf8} also shows the
significant difference between the cross sections calculated with
the approximation of Eq.~(\ref{msabit}) and the full result
obtained from Eq.~(\ref{tfi33}). This discrepancy is due to the
``factorization'' approximation, since the integrations of both
spin-independent and spin-dependent amplitudes over the internal
momentum weighted by the two wave functions lead to destructive
interference and further reduce the cross section. The cross
section for kaon photoproduction is in fact very small, of the
order of about 1 nb/sr at most, and even smaller for larger kaon
angles. The underlying reason is the lack of high momentum
components in the $^3_\Lambda{\rm H}$ wave function, inhibiting
hypernuclear formation. Nevertheless, the electromagnetic
production of the hypertriton has to be compared to the production
with strong probes, e.g. $d(p,K^+)^3_{\Lambda}{\rm H}$ whose cross
sections also have been predicted to be around 1
nb/sr\cite{komarov}. As shown in Fig.~\ref{hyplabf9}, $S$--waves
alone are insufficient to describe the reaction. Inclusion of the
higher partial waves further reduce the cross section by a factor
of more than three.  This can be traced to an overlap of $D$--wave
components in the $^3$He wave function with the dominant $S$--wave
in the hypertriton. In comparison, the higher partial waves in
pion photo- and electroproduction\cite{tiator2} decrease the cross
section by at most 20\%. Fig.~\ref{hyplabf9} also compares
calculations with $S$--waves using both a simple analytical model
for the hypertriton wave function \cite{congleton} and a
correlated three-body Faddeev wave function \cite{miyagawa2} that
includes the proper short-range behavior.  While the cross
sections obtained with the Faddeev wave functions does show more
structure, the difference in magnitude are only of order 10 -
20\%. Calculations coming up on $^4_\Lambda$He and $^4_\Lambda$H
\cite{hiyama2000} will supplement these studies  on $^3_\Lambda$H
and, because of their richer spectra, will allow an even closer
examination of the $YN$ force.

\section{\bf \normalsize QUASIFREE KAON PRODUCTION}

Due to the sizable momentum transfer to the hypernuclear system
the probability of forming such bound states is in fact rather
small. Ref.~\cite{cotanch86} has estimated this formation
probability to be around 5-10\% of the total $(\gamma, K^+)$
strength on nuclear targets, thus most of the kaon production
events will come from quasifree production, A$(\gamma, K Y)$B,
where the kaon can be a $K^+$ or $K^0$, and the hyperon can be
either a $\Lambda$ or a $\Sigma$.

\subsection{Quasifree kaon production on the deuteron}

In order to explore the $YN$ force more directly, hyperon
production processes on the deuteron, such as $\gamma( d,K^+)YN$,
appear as natural candidates. The hope is that the pole structure
of the $YN$ scattering operator will have visible effects in such
a production process. In a recent study \cite{tpole} the
$S$-matrix pole structure for the $YN$ system has been
investigated for various presently used $YN$ forces. As is well
known there is no bound state in the $\Lambda(\Sigma)N$ system,
but the present potential models support poles of the $S$-matrix
which are close to the $\Lambda$ and $\Sigma$ thresholds. Near the
$\Lambda$ threshold there are two $S$-wave virtual states at about
$-3$ and $-5$ MeV, and close to the $\Sigma$ threshold there is a
$^3{\rm S}_1\,$--$\,^3{\rm D}_1$ pole which appears at different
unphysical sheets of the Riemann energy surface depending on the
potential used.  This pole causes cusp-like structures in the
$\Lambda N$ scattering at the $\Sigma$ threshold. Their forms and
strengths depend on the potential employed.

Pioneering work in inclusive and exclusive $K^+$ photoproduction
on the deuteron has been done before \cite{wright} based on simple
hyperon-nucleon forces. These calculations suggested that
significant $YN$ final-state interaction effects be present near
the production thresholds.  Recently, these results were
reexamined \cite{yamamura99} using various recently formulated
$YN$ forces \cite{nijmegen1,nsc97} together with realistic $NN$
forces and an updated elementary photoproduction operator of the
$K^+Y$ pair on a nucleon.  For kaon photoproduction on the
deuteron the nuclear matrix element can be conveniently rewritten
by applying the M\"oller wave operator generating the final
scattering state
 to the right:
\begin{eqnarray}
\langle \,\Psi^{(-)}_{\bvec{q}_Y
\mu_Y\nu_Y\mu_N\nu_N}\,|\,t_{\gamma K}(1)\,|\,\Phi_d
\mu_d\,\rangle &\equiv&\langle \,\bvec{q}_Y
\mu_Y\nu_Y\mu_N\nu_N\,|\,T_Y\,|\,\Psi_d \mu_d\,\rangle \label{ic3}
\end{eqnarray}
Since we allow for $\Lambda -\Sigma$ conversion the state $\langle
\Psi^{(-)}_{\bvec{q}_Y\mu_Y\nu_Y\mu_N\nu_N}\,|$ as well as the
corresponding free state $\langle \bvec{q}_Y
\mu_Y\nu_Y\mu_N\nu_N\,|$ is a row with a $\Lambda$ and a $\Sigma$
component. The operator $T_Y$ applied to the deuteron state obeys
the integral equation

\begin{equation}
T_Y\,|\,\Psi_d \mu_d\,\rangle \,=\,t_{\gamma K}^Y (1)\,|\,\Psi_d
\mu_d\,\rangle \,+\, \sum_{Y^\prime} V_{Y,Y'} G_0^{Y'} T_{Y'}
\,|\,\Psi_d \mu_d\,\rangle \label{ic6}
\end{equation}
Equation~(\ref{ic6}) contains the elementary operator $t^Y_{\gamma
K}$ producing a specific hyperon $Y$ and $V_{Y,Y'}$ is the
hyperon-nucleon force including $\Lambda-\Sigma$ conversion.

\begin{figure}[t]
\begin{minipage}[htb]{75mm}
\centerline{\psfig{file=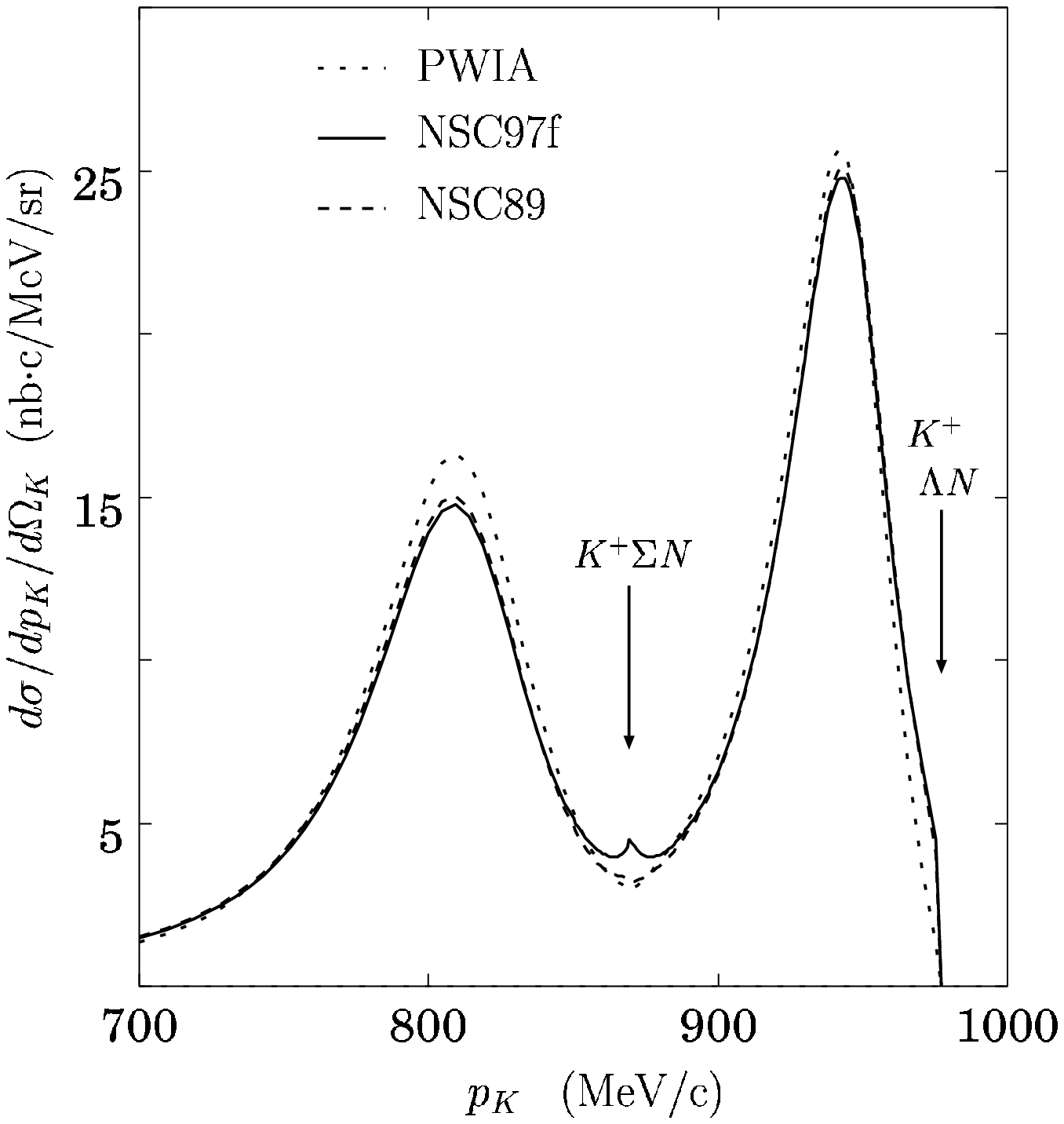,width=8cm}}
\end{minipage}
\hspace{\fill}
\begin{minipage}[ht]{75mm}
 \centerline{\psfig{file=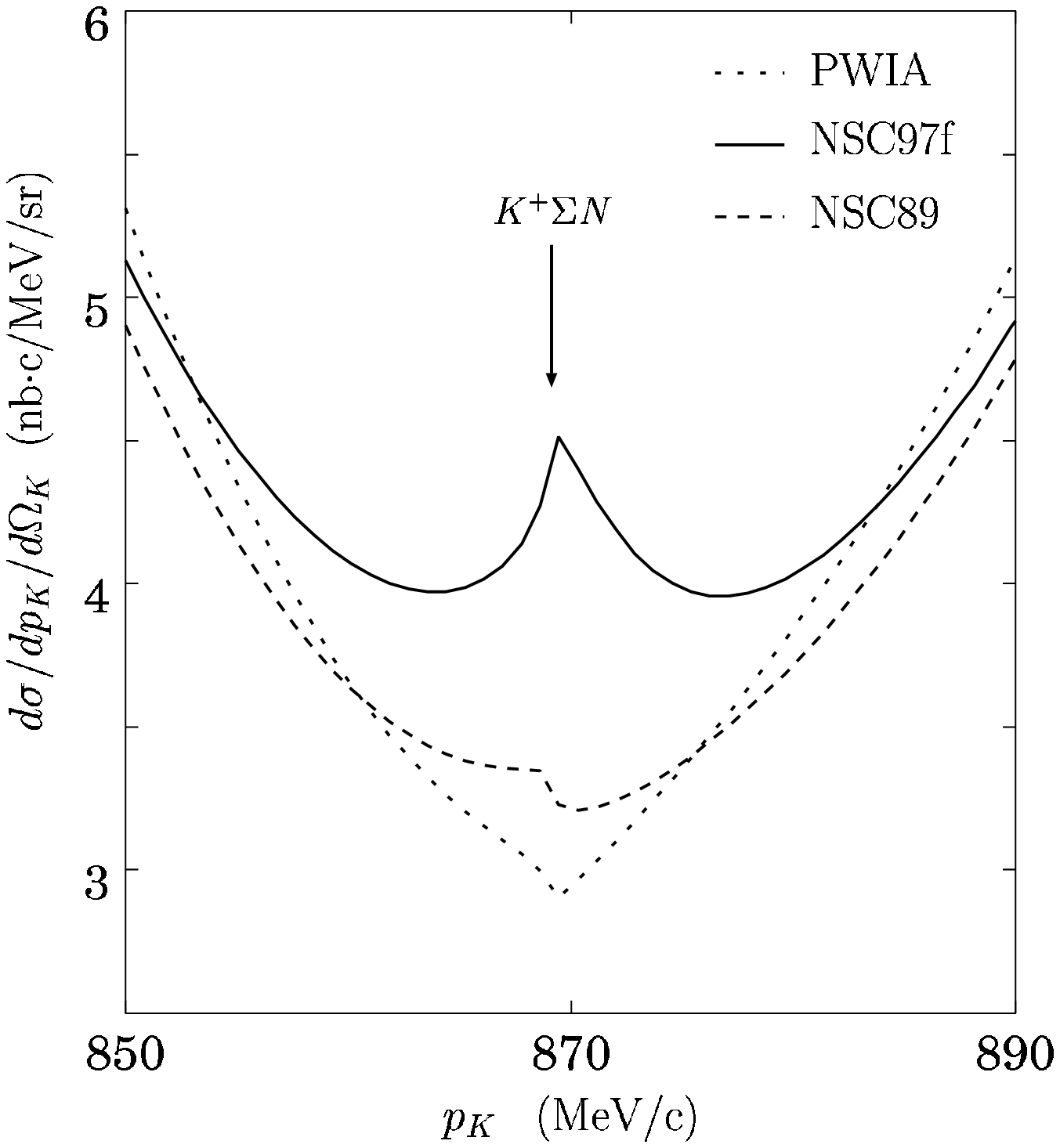,width=8cm}}
\end{minipage}
\caption{The left side shows the inclusive $\gamma( d,K^+)$ cross
section as a function of lab momenta $p_K$ for $\theta_K=0^\circ$
and photon lab energy $E_\gamma=1.3$ GeV.
 The plane wave result is compared to two $YN$ force predictions.
The FSI effects are especially pronounced near the $K^+\Lambda N$
and $K^+\Sigma N$ thresholds (see the enlarged figure on the right
side), the locations of which are indicated by the arrows. The
right side shows the results enlarged around the $K^+\Sigma N$
threshold. The figures are taken from Ref.~\cite{yamamura99}.}
\label{hyplabf10}
\end{figure}

Figure \ref{hyplabf10} compares inclusive cross sections for $d
(\gamma , K^+)$ in plane wave impulse approximation (PWIA) with
calculations that include FSI generated with the hyperon-nucleon
forces NSC89 \cite{nijmegen1} and NSC97f \cite{nsc97} which both
lead to the correct hypertriton binding energy. The two pronounced
peaks around $p_K=$ 945 and 809 MeV/c can be understood in PWIA.
They are due to quasi-free processes, where one of the nucleons in
the deuteron is a spectator and has zero momentum in the lab
system. This then leads to a vanishing argument $q=0$ in the
deuteron wavefunction, which causes the peaks. Under this
condition the kinematics of the $\gamma$-induced process on a
single nucleon fixes the peak positions for $p_K$ in the lab
system. Figure~\ref{hyplabf10} shows significant deviations
between the plane wave result and the results with FSI based on
the NSC89 and NSC97f hyperon-nucleon forces. Near the $K^+\Lambda
N$ threshold the FSI enhances the cross section by up to 90\%.
Near the $K^+\Sigma N$ threshold the effects are even more
dramatic. While NSC89 slightly enhances the cross section, NSC97f
leads to a much stronger effect with a prominent cusp-like
structure.  The two $YN$ potentials lead to predictions which
differ by up to 35\%. This can be traced back to the location of
the $S$-matrix pole for the $\Lambda N-\Sigma N$ system around the
$\Sigma N$ threshold. Each of the two $YN$ potentials generate a
pole in the state $^3{\rm S}_1\,$--$\,^3{\rm D}_1$ near $p_{\Sigma
N}=0$. The potential NSC89 leads to a pole position which in a
single channel case would be called a virtual state (in this case
it would lie exactly on the imaginary axis). The coupling of the
$\Lambda$ and $\Sigma$ channels moves the pole for the NSC97f
force away from the positive imaginary axis into the second
$p_{\Sigma N}$ quadrant. In a time-dependent description the
energy related to that pole position leads to a decreasing
amplitude. In the literature, this sort of pole is sometimes
referred to as an `unstable bound state'. Apparently, the actual
pole position depends on the details of the $YN$ force. The pole
positions are an inherent property of the $YN$ forces and the
actual location chosen by nature should be determined with the
help of experimental measurements.

\subsection{Quasifree kaon production on heavier nuclei}

Quasifree production on heavier nuclei allows for the study of the
reaction process in the nuclear medium as well as final state
interaction effects without being obscured by the details of the
nuclear transitions as discussed above. The predictions presented
here \cite{lee99} are in a DWIA framework that has been
successfully applied in previous work on quasifree pion photo- and
electroproduction \cite{lee93} and eta photoproduction on nuclei
\cite{lee96}. The key ingredients are:
\begin{enumerate}
\item the single-particle wave function and spectroscopic factor,
usually taken from electron scattering,
\item the elementary kaon photoproduction amplitude,
\item the distorted kaon wave function which can be taken from kaon
elastic scattering in case of the $K^+$,
\item the hyperon-nucleus final-state interaction.
\end{enumerate}

In contrast to hypernuclear production discussed above, the
reaction is {\em quasifree}, meaning that the magnitude of
$Q$ has a wide range, including zero. Since the reaction
amplitude is proportional to the Fourier transform of the bound
state single particle wavefunction, it falls off quickly as the
momentum transfer increases. Thus, we will restrict ourselves to
the low $Q$ region ($<$ 500 MeV/c) where the nuclear recoil
effects can be safely neglected for nuclei of $A > 6$.  The
nuclear structure aspects are now contained in the spectroscopic
factor, and in the single-particle matrix element of
Eq.~(\ref{sixdim}) the bound-state $\Lambda$ wave function has to be
replaced by a scattering state, obtained from solving a
Schrodinger equation with some optical potential.

\subsection{Hyperon-nucleus optical potentials}

\begin{figure}[!b]
\centerline{\psfig{file=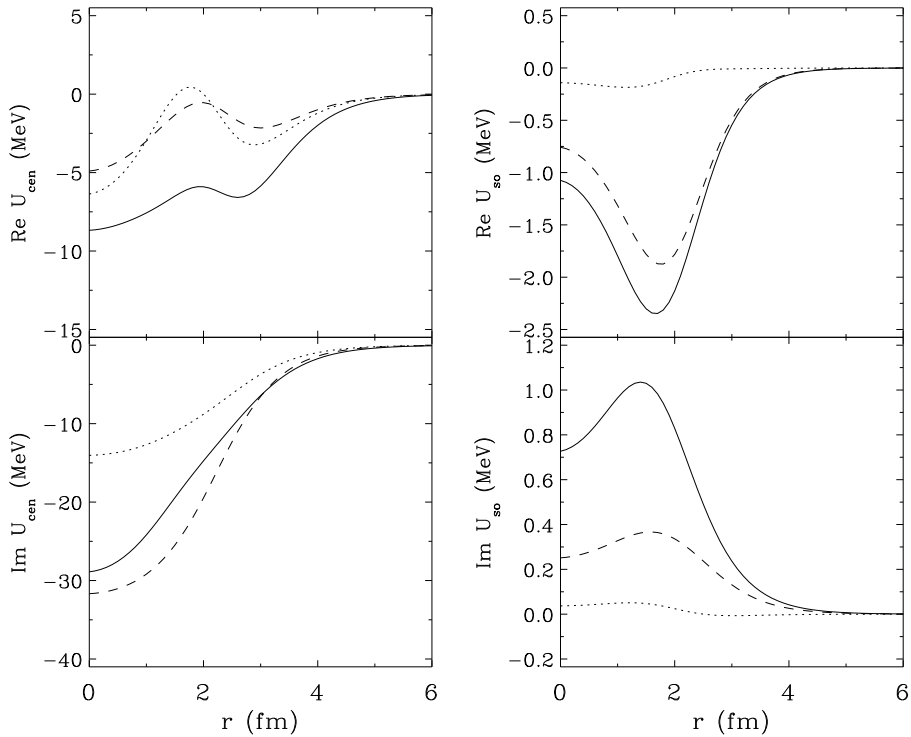,width=12cm}} \caption{Hyperon
optical potentials for $^{12}$C at 200 MeV kinetic energy. The
dotted line shows the $\Lambda$ potential, while the dashed line
depicts the $\Sigma^0$ potential. The proton potential (solid
line) is shown for comparison.} \label{hyplabf11}
\end{figure}

Very few optical potentials have been constructed for the
$\Lambda$ and $\Sigma$, mostly due to lack of data. Here, we
employ the global optical model of Ref.~\cite{cooper94}. It is
based on a global nucleon-nucleus Dirac optical potential fit
\cite{cooper93}. The parameters of the potential are motivated by
the constituent quark model and adjusted to fit the hypernuclear
binding energy data.  We use its nonrelativistic equivalent
version which has a central and a spin-orbit part, $U(r)=U_{\rm
cen}(r)+U_{\rm so}(r)\; \bvec{\sigma} \cdot \bvec{l}$. Note that
the spin-orbit part is multiplied by a factor that depends on the
partial wave under consideration. Fig.~\ref{hyplabf11} shows the
real and imaginary parts of both $U_{\rm cen}(r)$ and $U_{\rm
so}(r)$ on $^{12}$C at 200 MeV kinetic energy for the $\Lambda$
and the $\Sigma^0$. For comparison, they are also shown for the
proton. The real parts of the central potential are clearly
smaller than the proton potential by around a factor of two,
reflecting the fact that Lambdas and Sigmas have a smaller binding
energy in hypernuclei. The imaginary part of the $\Sigma$'s
central potential is similar in magnitude to that of the nucleon,
due to the large $\Sigma N \rightarrow \Lambda N$ conversion
width. The very small spin-orbit potential of the $\Lambda$ is a
reflection of the $\Lambda N$ spin-orbit force which is known to
be small.

\begin{figure}[!t]
\centerline{\psfig{file=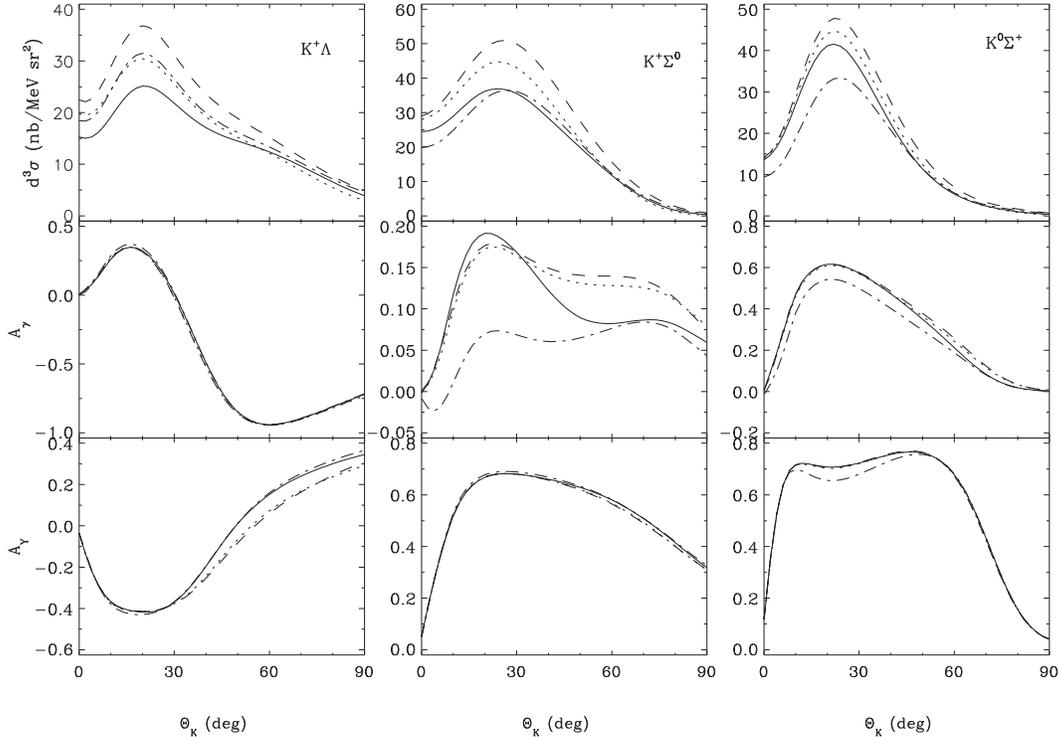,width=14cm}}
\caption{Results for the differential cross section, the polarized
photon asymmetry, and the hyperon recoil polarization for the
reaction $^{12}{\rm C}(\gamma,KY){\rm B_{g.s.}}$ at $E_\gamma$=1.4
GeV and $Q=120$ MeV/c under quasifree kinematics. Three of the six
possible channels are shown in the three columns. The four curves
correspond to calculations in PWIA (dashed), in DWIA with kaon
only distorted (dotted), with hyperon only distorted
(dash-dotted), and with both distorted (solid).} \label{hyplabf12}
\end{figure}

Results are shown in Fig.~\ref{hyplabf12} for the reaction
$^{12}{\rm C}(\gamma,KY) ^{11}{\rm B_{g.s.}}$ at $E_\gamma$=1.4
GeV and $Q=120$ MeV/c under quasifree kinematics. As the kaon angle
increases, the kaon energy decreases while the hyperon energy
increases. In this particular case, the kaon and hyperon energies
can reach around 500 to 700 MeV.  Figure~\ref{hyplabf12} shows the
differential cross sections as well as two polarization
observables, comparing PWIA calculations with results that include
hyperon and kaon final state interaction (FSI). Kaon distortion
reduces the cross sections by about 10-20\% but has little effect
on the polarization observables.  Including the hyperon FSI
reduces the angular distributions  by up to 30\% at forward
angles.  Again, with the exception of the photon polarization in
$K^+ \Sigma^0$ production, the polarization observables are barely
affected by the inclusion of FSI.  This situation is similar to
previous findings in quasifree pion and eta photoproduction
\cite{lee93,lee96}. It opens the possibility to use the
polarization observables as a way to study modifications of the
basic production process in the nuclear medium. The magnitudes of
the $\Sigma$ and $P$ observables is sizeable and should
 be measurable.

\section{\bf \normalsize CONCLUSIONS}

Driven by new experimental results from facilities like Jlab and
ELSA, the field of Strangeness production with photons and
electrons is experiencing a revival.  Experimental proposals, some
approved at Jlab over ten years ago, are being carried out and
begin to produce data of unprecedented quality.  In the meantime,
our understanding of the elementary production mechanism has
evolved significantly over the last decade.  Within effective
Lagrangian approaches the need for hadronic form factors has
become clear, along with their implications for gauge invariance.
Born coupling constants, long found to be too small in model fits,
can now be reconciled with their SU(3) values and upcoming
high-quality data should allow extracting their precise values by
extrapolating to a Born pole using dispersion relations.  In the
resonance sector new coupled-channels analyses are beginning to
shed light on the relevant resonances by combining data from kaon
photoproduction with information from hadronic strangeness
production and with the vast reservoir of data in pion scattering
and pion photoproduction.  Attempts to describe threshold
production through Chiral Perturbation Theory are still in their
infancy but may eventually lead to a better understanding of the
validity of chiral expansions in the SU(3) sector.  With our
knowledge of the elementary kaon production amplitude expanding,
predictions for kaon productions on nuclear targets are becoming
more reliable. Here the few-body sector is especially appealing
due to the rigor of the theoretical calculations. Modern
hyperon-nucleon forces can be studied directly in quasifree kaon
production on the deuteron. The $A=3$ and 4 hypernuclei can now be
handled theoretically without approximations and, using the
nucleus as a spin-isospin filter, place severe constraints on
certain parts of the $YN$ interaction.  Jlab experiments are just
beginning to take advantage of the photoproduction mechanism to
excite unnatural parity and high-spin states; this spectroscopic
information will supplement our knowledge of the effective
$\Lambda N$ interaction from present hadronic probes employed at
KEK and BNL. The full power of the almost distortionless
$(\gamma,K^+)$ reaction to populate deeply bound states in heavier
nuclei and study details of the $\Lambda$ wave function will
hopefully be used in second-generation experiments at Jlab and
MAMI C.  In the coming years, these experimental and theoretical
developments should lead to a much improved understanding not only
of the $YN$-force, but also of hadronic physics in the SU(3) realm
in general.

\section*{\bf \normalsize Acknowledgements}
C.B., H.H., F.X.L., and L.E.W. acknowledge the support from the US
Department of Energy. W.G. and L.T. would like to thank the
Deutsche Forschungsgemeinschaft for financial support. S.S.K.
thanks the Heisenberg-Landau program for support. T.M.
acknowledges support from the Deutscher Akademischer
Austauschdienst (DAAD). C.B., T.M. and S.S.K. are grateful for the
hospitality of the Insitut f\"ur Kernphysik at the University of
Mainz.

\end{document}